\documentclass[aps,floatfix,twocolumn]{revtex4}
\usepackage{amsmath,amssymb}
\usepackage{cancel}
\usepackage{natbib}
\usepackage{color}
\newcommand{\mc}[1]{\mbox{$\mathcal{#1}$}}

\begin{document}

%
\title{Generalized Swiss-Cheese Cosmologies II: Spherical Dust}
\author{C\'{e}dric Grenon}
\email{cgrenon@astro.queensu.ca}
\author{Kayll Lake}
\email{lake@astro.queensu.ca}
\affiliation{Department of Physics, Queen's University, Kingston,
Ontario, Canada, K7L 3N6 }
\date{\today}
\begin{abstract}
The generalized Swiss - cheese model, consisting of a Lema\^itre - Tolman (inhomogeneous dust) region matched, by way of a comoving boundary surface, onto a Robertson-Walker background of homogeneous dust, has become a standard construction in modern cosmology. Here we ask if this construction can be made more realistic by introducing some evolution of the boundary surface. The answer we find is no. To maintain a boundary surface using the Darmois - Israel junction conditions, as opposed to the introduction of a surface layer, the boundary must remain exactly comoving. The options are to drop the assumption of dust or allow the development of surface layers. Either option fundamentally changes the original construction.
\end{abstract}
\maketitle
%
%
\section{Introduction}
In \cite{cedric} we generalized the Swiss-cheese cosmologies so as to include nonzero linear momenta of the associated boundary surfaces. We found that the final effective gravitational mass and size of the evolving inhomogeneities depends on their linear momenta but these properties are essentially unaffected by the details of the background model. In this paper we explicitly specify the inhomogeneity and take it to be Lema\^itre - Tolman (L - T) inhomogeneous dust. With a comoving boundary surface, this construction is a standard feature in modern cosmology \cite{Bolejko} \cite{Bolejko2011a}. Since an exactly comoving boundary surface is clearly a mathematical idealization, the purpose of this paper is to examine possible evolution of the boundary surface. We find that if we rigorously maintain the Darmois - Israel junction conditions, no evolution of the boundary surface is possible. Hints that this standard construction is a mathematical artifact are reviewed in \cite{Krasinski1}. Evolution away from dust, or the development of surface layers (which we briefly discuss) fundamentally changes the original construction.
%
%
\section{Lema\^{i}tre-Tolman Solutions in Lagrangian Form}
We find it convenient to use the somewhat less common Lagrangian coordinates to describe the L - T solution and we begin by summarizing some fundamental properties of the solution in these coordinates.
\subsection{Basic calculations}\label{basic}
Starting with Einstein's equations \cite{notation}
\begin{equation}\label{einstein}
    G_{\alpha \beta}+\Lambda g_{\alpha \beta}=8 \pi T_{\alpha \beta},
\end{equation}
where $\Lambda$ is the cosmological constant, we seek solutions for dust,
\begin{equation}\label{dust}
    T_{\alpha \beta}=\rho \; u_{\alpha}u_{\beta}
\end{equation}
where the $u^{\alpha}$ are tangent to the generators of the geodesic flow and we consider only positive definite energy densities $\rho>0$. We use comoving synchronous coordinates so that
\begin{equation}\label{metric}
ds^2=e^{\alpha(m,t)}dm^2+R^2(m,t)d\Omega^2-dt^2
\end{equation}
where $d\Omega^2$ is the metric of a unit two-sphere, which we write in the usual form $d\theta^2+\sin^2(\theta)d\phi^2$, and we assume the existence of  an origin defined by $R(0,t)=0$. The generators of the flow are then $u^{\alpha}=\delta^{\alpha}_{t}$ and the radial normals are $n^{\alpha}=\pm e^{-\alpha/2} \delta^{\alpha}_{m}$ so that $-u^{\alpha}u_{\alpha}=n^{\alpha}n_{\alpha}=1$ and $u^{\alpha}n_{\alpha}=0$. From
\begin{equation}\label{Gun}
    G_{\alpha \beta}u^{\alpha}n^{\beta}=0
\end{equation}
we find
\begin{equation}\label{first}
    e^{\alpha}=\frac{(R^{'})^2}{1+2E(m)}
\end{equation}
where $E$ is an arbitrary function ($>-1/2$). Define
\begin{equation}\label{emass}
\mathcal{M}(m,t) \equiv \frac{R^3}{2}\mathcal{R}_{\theta \phi}^{\;\;\;\;\theta \phi},
\end{equation}
where $\mathcal{R}$ is the Riemann tensor and so $\mathcal{M}$ is the effective gravitational mass \cite{cedric}.
We obtain
\begin{equation}\label{rdot}
   \dot{R}^2=2E+\frac{2\mathcal{M}}{R}.
\end{equation}
With the gauge condition \cite{gauge}
\begin{equation}\label{guage}
   \mathcal{M}(m,t)=m+\Lambda \frac{R^3}{6}
\end{equation}
the information required to solve Einstein's equations is simply (\ref{rdot}), the $m$ derivative of (\ref{rdot}) and equation (\ref{guage}). We arrive at
\begin{equation}\label{rho}
    4 \pi \rho(m,t) = \frac{1}{R^2R^{'}} .
\end{equation}
The implicit solution to (\ref{rdot}) is
\begin{equation}\label{tb}
    t-t_{B}(m)=\pm \int_{0}^{R} \frac{dx}{\sqrt{2E+\frac{\Lambda x^2}{3}+\frac{2m}{x}}}
\end{equation}
where $t_{B}$ is a second arbitrary function.
%
%
\subsection{Invariants}
The L - T solutions have two independent invariants derivable from the Riemann tensor without differentiation. These can be taken to be \cite{Santosuosso}
\begin{equation}\label{Ricci}
    \mathcal{R}=2(4 \pi \rho + 2 \Lambda)
\end{equation}
and
\begin{equation}\label{Weyl}
   w=\frac{16}{3}\left(4 \pi \rho-\frac{3m}{R^3}\right)^2
\end{equation}
where $\mathcal{R}$ is the Ricci scalar and $w$ is the first Weyl invariant ($C_{\alpha \beta \gamma \delta}C^{\alpha \beta \gamma \delta}$ where $C_{\alpha \beta \gamma \delta}$ is the Weyl tensor).
%
%
\subsection{The Origin}
From (\ref{rho}) we have
\begin{equation}\label{limit}
    \lim_{m \rightarrow 0}\frac{R^3}{m}=\lim_{m \rightarrow 0}3R^2R^{'}=\frac{3}{4 \pi \rho(0,t)},
\end{equation}
and so from (\ref{Weyl})
\begin{equation}\label{weylc}
    w(0,t)=0.
\end{equation}
Further, it follows immediately from (\ref{rdot}) that
\begin{equation}\label{ecenter}
    \lim_{m \rightarrow 0}E=0.
\end{equation}
%
%
\subsection{Singularities}
It is clear from (\ref{rho}) and (\ref{Ricci}) that scalar polynomial singularities occur for
\begin{equation}\label{sing}
    R^2R^{'}=0
\end{equation}
where $\rho$, and therefore $\mathcal{R}$ and $w$, diverge. A ``bang" (or ``crunch") occurs for $R=0$.
Shell crossing singularities occur for $R^{'}=0$. The conditions for their avoidance are well known for $\Lambda=0$ \cite{Hellaby1} and these conditions have recently been extended to the case $\Lambda \neq 0$ \cite{Wainwright}.
%
%
\subsection{Conformal Flatness}\label{flatness}
Conformal flatness of the L - T solutions can be recognized by the following necessary and sufficient condition for conformal flatness \cite{lake}: $\Delta=0$ where
\begin{equation}\label{cf1}
    \Delta \equiv m-\frac{4}{3} \pi R^3 \rho.
\end{equation}
From (\ref{cf1}) and (\ref{rho}) then
\begin{equation}\label{cf2}
   R^{'}=\frac{R}{3m}
\end{equation}
so that
\begin{equation}\label{cf3}
    R=a(t)m^{1/3}.
\end{equation}
Using (\ref{cf3}) in (\ref{rdot}) with (\ref{guage}) we find
\begin{equation}\label{erw}
   \frac{2E}{m^{2/3}}=\dot{a}^2-\frac{2}{a}-\frac{a^2 \Lambda}{3}
\end{equation}
and so both sides of this relation have to be constants (the left being independent of $t$ and the right being independent of $m$). The constancy of the term on the right is, of course,  the Friedmann equation for dust.
Finally, if $a(t=c)=0$ it follows from (\ref{tb}) that $t_{B}(m)=c$ where $c$ is a constant. In summary, the condition $\Delta=0$ reduces (\ref{metric}) to the Robertson - Walker (R - W) metric.
%
%
\section{Junction Conditions}\label{junctionconditions}
The theory of hypersurfaces ($\Sigma$, here considered to be timelike) in spacetime is well established (see, for example, \cite{poisson}) but we think it important to cover a few details here. We follow the notation of \cite{musgrave}. The Darmois \cite{Darmois} - Israel \cite{Israel} junction conditions for boundary surfaces require the continuity of the first and second fundamental forms accross $\Sigma$ \cite{eisenhart}. These conditions are in general not equivalent to the conditions of Lichnerowicz \cite{Lich} which require the continuity of the metric and all first order partial derivatives of the metric across $\Sigma$ in coordinates that traverse $\Sigma$ (such coordinates being referred to as ``admissible"). The only case when the two sets of conditions are equivalent is when Gaussian - normal coordinates \cite{MTW} are used. For example,  taking $(m,\theta,\phi,t)$ continuous through $\Sigma$, and using the Lichnerowicz conditions, we arrive at the constraint $[\rho]=0$ even for $dm/d \tau = 0$ along $\Sigma$. This constraint is erroneous as we discuss below. Whereas one might argue that the coordinates $(m,\theta,\phi,t)$ are not ``admissible", here, to avoid any source of confusion, we rely only on the Darmois - Israel conditions.

The basic model consists of randomly distributed non-intersecting timelike spherical boundary surfaces $\Sigma$ in a R - W background $\mathcal{V}^{+}$. The local inhomogeneities $\mathcal{V}^{-}$ are L - T solutions. Both $\mc{V}^\pm$ have metrics of the general form (\ref{metric}) with $\mc{V}^+$ being conformally flat. On $\Sigma$ we write
\begin{equation}\label{intrinsicg}
    ds^2_{\Sigma}=g_{i j}d x^{i} d x^{j}
\end{equation}
where $g_{i j}$ is the metric intrinsic to $\Sigma$ (first fundamental form) and $x^{i}$ are the coordinates intrinsic to $\Sigma$. With spherical symmetry, without loss in generality we take the coordinates $\theta$ and $\phi$ (but not $t$) continuous through $\Sigma$ on which we write the intrinsic metric as
\begin{equation}\label{intrinsic}
    ds^2_{\Sigma}=R^2(\tau)d\Omega^2-d \tau^2
\end{equation}
where $\tau$ is the proper time on $\Sigma$. Note that the proper time  derivative $d/d \tau$ is defined only along $\Sigma$.

Now our use of the Lagrangian variable $m$ requires special attention. A straight - forward calculation gives
\begin{equation}\label{Kthetatheta}
        K^2_{\theta \theta}=R^2\left(\left(\frac{dR}{d \tau}\right)^2+1-\frac{2 \mathcal{M}}{R}\right),
\end{equation}
where $K_{i j }$ is the curvature extrinsic to $\Sigma$ (second fundamental form).
For boundary surfaces, if follows from (\ref{Kthetatheta}) that $[\mathcal{M}]=0$. Throughout we consider $\Lambda$ a universal constant and set $[\Lambda]=0$. It now follows from (\ref{guage}) that
\begin{equation}\label{masscont}
    [m]=0.
\end{equation}

We write $\pm \tilde{n}^{\alpha}|^{\pm}_{\Sigma}$ as the unit radial 4 - orthogonal to $\Sigma$ and $\tilde{u}^{\alpha}|^{\pm}_{\Sigma}$ as the unit 4 - tangent to $\Sigma$. We simply drop $|^{\pm}_{\Sigma}$ and write
\begin{equation}
\tilde{n}^\alpha=\left(e^{-\alpha/2}\frac{dt}{d\tau},0,0,e^{\alpha/2}\frac{dm}{d\tau}\right),
\end{equation}
and
\begin{equation}
\tilde{u}^\alpha=\left(\frac{dm}{d\tau},0,0,\frac{dt}{d\tau}\right)\,.
\end{equation}
The vectors $\tilde{n}^\alpha$ and $\tilde{u}^\alpha$ are unit vectors in consequence of the relation
\begin{equation}\label{unit}
    e^\alpha\left(\frac{dm}{d \tau}\right)^2-\left(\frac{dt}{d \tau}\right)^2=-1
\end{equation}
which follows from the fact that $\tau$ is the proper time on $\Sigma$.
We choose the global time-orientation $dt/d \tau>0$. The use of the Lagrangian variable $m$ makes the rigging of $\tilde{n}^\alpha$ straightforward as $m$ increases monotonically away from the origin. When $\Sigma$ is comoving $dm/d\tau=0$ and we get $\tilde{n}^\alpha=n^\alpha$ and $\tilde{u}^\alpha=u^\alpha$, the normals and generators of the fluid flow.
%
%
\subsection{Boundary surfaces}\label{boundarysurfaces}
Let us assume that $\Sigma$ is a boundary surface. Then, by assumption, the following well - known necessary (but not sufficient) conditions hold:
\begin{equation}
\label{eqn-Gnn1}
\left[ G_{\alpha \beta} \tilde{n}^\alpha \tilde{n}^\beta \right] = \left[ G_{\alpha \beta}
 \tilde{u}^{\alpha}
   \tilde{n}^\beta \right]= 0.
\end{equation}
From Einstein's equations then we have
\begin{equation}\label{ab}
   [A]=[B]=0
\end{equation}
where
\begin{equation}\label{A}
    A \equiv T_{\alpha \beta}\tilde{n}^{\alpha}\tilde{n}^{\beta}
\end{equation}
and
\begin{equation}\label{B}
    B \equiv T_{\alpha \beta}\tilde{u}^{\alpha}\tilde{n}^{\beta}.
\end{equation}
From (\ref{einstein}), (\ref{dust}) and (\ref{metric}) we find
\begin{equation}\label{Aa}
   A=e^{\alpha} \rho \left(\frac{dm}{d \tau}\right)^2
\end{equation}
and
\begin{equation}\label{Bb}
   B=e^{\alpha/2} \rho \frac{dm}{d \tau}\frac{dt}{d \tau}.
\end{equation}
The standard assumption is to set $\Sigma$ comoving so that $dm/d \tau = 0$ and as a result the conditions (\ref{ab}) are empty in the sense that $A=B=0$. However, in this case, in addition to (\ref{masscont}), one very frequently encounters the additional constraint $[\rho]=0$ (which, as explained above, follows from the Lichnerowicz conditions) so that from (\ref{cf1}) we have $[\Delta]=0$ and so $\Delta=0$ along $\Sigma$. To see that this additional constraint is spurious, one need only reflect upon classics: the Einstein - Straus model or Oppenheimer - Snyder collapse!

\subsection{Non-comoving boundary surfaces: $dm/d \tau|_{\Sigma} \neq 0$}\label{noncomoving}
From (\ref{Aa}) and (\ref{Bb}) we have
\begin{equation}\label{jumpab}
    0=\left[\frac{B^2}{B^2-A^2}\right]=\left[\left(\frac{dt}{d \tau}\right)^2\right].
\end{equation}
From (\ref{jumpab}) and $[\tilde{u}^{\alpha}\tilde{u}_{\alpha}]=0$ then
\begin{equation}\label{jumpmdot}
    \left[e^{\alpha}\left(\frac{dm}{d \tau}\right)^2\right]=0
\end{equation}
so that from (\ref{Aa})
\begin{equation}\label{jumprho}
    [\rho]=0.
\end{equation}
Condition (\ref{jumprho}) is, for $dm/d \tau|_{\Sigma} \neq 0$, not an additional constraint, but rather a requirement of the Darmois - Israel junction conditions.
In summary, with (\ref{jumprho}), for a boundary surface with $d m/d \tau|_{\Sigma} \neq 0$, and (\ref{cf1}) we have shown that
\begin{equation}\label{conditions}
    [\Delta]=0.
\end{equation}
This tells us that if $\Sigma$ is a boundary surface and $dm/d \tau|_{\Sigma} \neq 0$ along $\Sigma$, then $\mathcal{V}^{-}$ is conformally flat at $\Sigma$. Now since $dm/d \tau|_{\Sigma} \neq 0$ we can consider a region $\mathcal{J}$ in $\mathcal{V}^{-}$ defined by the non-vanishing ranges $\delta m$ and $\delta t$ bounded by $\Sigma$, to the future of $\Sigma$ for $dm/d \tau|_{\Sigma} > 0$ and to the past of $\Sigma$ for $dm/d \tau|_{\Sigma} < 0$. Since the unique evolution of an R - W flow is the same R - W flow, the R - W solution in $\mathcal{V}^{-}$ is the same as that in $\mathcal{V}^{+}$. There cannot then exist non-comoving boundary surfaces in spherical dust that signify local inhomogeneities.
%
%
\subsection{Surface layers}\label{thinshell}
The Israel formulation of thin shells (which we discuss only briefly here for completeness) involves the construction of surface properties of $\Sigma$ out of discontinuities in the extrinsic curvature, by way of the Lanczos equations \cite{poisson} \cite{musgrave} \cite{Israel}. In particular, the surface energy density of $\Sigma$ is given by
\begin{equation}\label{surfaceenrgy}
    8 \pi \sigma = -\gamma-\gamma_{i j}u^iu^j,
\end{equation}
where $\gamma _{i j} \equiv [K_{i j}]$, $\gamma \equiv \gamma^i_i$ and $u^i$ ($\equiv d x^i/d \tau$) is the three - tangent to $\Sigma$. The surface pressure ($p$) is defined by
\begin{equation}\label{surfacepressure}
    16 \pi p = \gamma -\gamma_{i j}u^iu^j.
\end{equation}
These phenomenological properties of $\Sigma$ are related to the evolution of $\Sigma$ via
\begin{equation}\label{psigma1}
\left[ \tilde{n}_\alpha \tilde{u}^\beta \nabla_\beta \tilde{u}^\alpha \right] = -\gamma_{i j}u^iu^j=8 \pi ( p + \sigma /2),
\end{equation}
and to the enveloping 4 - geometries via
\begin{equation}
\label{eqn-conservation1}
u^i \nabla_i \sigma + (\sigma + p)\nabla_i u^i =
  \left[ B \right]
\end{equation}
and
\begin{equation}
\label{eqn-lawN1}
-\-(\sigma + p) \overline{\tilde{n}_\alpha \tilde{u}^\beta \nabla_\beta \tilde{u}^\alpha} + p \overline{K^i_i}
= \left[ A \right].
\end{equation}
For the cases under consideration here we continue to use the Lagrangian variable $m$ and continue to classify the motion via $dm/d \tau |_{\Sigma}$.

For shells when we write down the standard equation for $\Sigma$, $dm/d \tau |_{\Sigma} = 0$, we must keep in mind that $[m]$ is not necessarily $0$. By writing down $dm/d \tau |_{\Sigma} = 0$ we mean $dm/d \tau |_{\Sigma}^{-} = dm/d \tau |_{\Sigma}^{+}=0$ (and so $[m]$ is constant along $\Sigma$).
It then follows immediately form (\ref{metric}) that $\Sigma$ is geodesic in $\mathcal{V}^{\pm}$. As a result, it follows from (\ref{psigma1}) that
\begin{equation}\label{psigmacon}
    p+\frac{\sigma}{2}=0.
\end{equation}
At first glance, the possible cases are: $p=\sigma=0$, the boundary surface discussed above, and the thin shell with surface equation of state $p=-\sigma/2$. However, from (\ref{dust}) and (\ref{eqn-lawN1}) we have
\begin{equation}\label{coshell}
    p\overline{K^{\theta}_{\theta}}=0
\end{equation}
and so either $p=0$ (again the boundary surface) or $K^{\theta}_{\theta}$ changes sign across $\Sigma$. Since
\begin{equation}\label{kthetatheta}
    K^{\theta}_{\theta}=\frac{R^3R^{'}}{e^{\alpha/2}}
\end{equation}
 it follows that $R^{'}$, and from (\ref{rho}) also $\rho$, must change sign across $\Sigma$. We reject this possibility and conclude that $\Sigma$ cannot represent a comoving thin shell.

Now for  $dm/d \tau |_{\Sigma} \neq 0$ we have two possibilities: (i) $dm/d \tau |_{\Sigma}^{-} = dm/d \tau |_{\Sigma}^{+}\neq0$ (and so $[m]$ remains constant along $\Sigma$) and (ii) $dm/d \tau |_{\Sigma}^{-} \neq dm/d \tau |_{\Sigma}^{+}$ (with at least one $\neq0$). Clearly the notation ($dm/d \tau |_{\Sigma}$) is inadequate for the second case. There is now a plethora of possibilities to consider for both cases, some of which have been examined in detail (see, for example, \cite{Krasinski1} and references therein). Basically, we can choose some evolution for $\Sigma$ and see what surface properties ($\sigma$ and $p$) result, or choose these surface properties and see what evolution of $\Sigma$ results. In both cases, the assumption that we have enveloping dust enters by way of the constraint equations (\ref{eqn-conservation1}) and (\ref{eqn-lawN1}). To represent an inhomogeneous cosmological model, as opposed to a purely mathematical construction, physical input for $\sigma$ and $p$, or the evolution of $\Sigma$, is needed. In any event, we are now quite some ways away from the generalized Swiss - cheese model.
%
%
\section{conclusion}
We have examined the generalized Swiss - cheese model, consisting of a Lema\^itre - Tolman (inhomogeneous dust) region matched, by way of a boundary surface, onto a Robertson-Walker background of homogeneous dust. A standard construction in modern cosmology is to assume that the boundary surface is exactly comoving. Here we asked if this construction can be made more realistic by introducing some evolution of the boundary surface. The answer we have found is no. We have found that to maintain a boundary surface using the Darmois - Israel junction conditions, as opposed to the introduction of a surface layer, the boundary must remain exactly comoving. We conclude that this standard construction is a mathematical artifact, not an acceptable physical model of a cosmological inhomogeneity. The options are to drop the assumption of dust or allow the development of surface layers. Either option fundamentally changes the original construction.

In closing we should note that the situation we have considered, though common, is very idealized in the sense that the background has been taken to be exactly Robertson - Walker. If this is not the case, $\Sigma$ can indeed evolve. Further discussion of various scenarios which have been studied can be found in \cite{Krasinski1} and \cite{gauge}.
%
\section*{Acknowledgments}
KL is supported by a grant from the Natural Sciences and Engineering Research Council of Canada. Portions of this work were made possible by use of \emph{GRTensorII}\cite{GRTensorII}.
%
%


\begin{thebibliography}{}
\bibitem{notation}We use geometrical units, a signature of $+2$ and designate functional dependence usually only on the first appearance of a function. Throughout, $^{'}=\partial / \partial m$, $^{.}=\partial/ \partial t$, $ [ \Psi ] \equiv \Psi^{+} - \Psi^{-}$ and $\bar{\Psi}\equiv (\Psi^{+} + \Psi^{-})/2$. Proper time derivatives along $\Sigma$ are written out explicitly, for example, $dR /d \tau$.
\end{thebibliography}
\end{document}